\documentclass[preprint]{vgtc}               

\ifpdf
  \pdfoutput=1\relax                   
  \usepackage{graphicx}                
  \DeclareGraphicsExtensions{.pdf,.png,.jpg,.jpeg} 
\else
  \usepackage{graphicx}                
  \DeclareGraphicsExtensions{.eps}     
\fi%

\graphicspath{{figures/}{pictures/}{images/}{./}} 

\usepackage{microtype}                 
\PassOptionsToPackage{warn}{textcomp}  
\usepackage{textcomp}                  
\usepackage{mathptmx}                  
\usepackage{times}                     
\usepackage{cite}                      
\usepackage{tabu}                      
\usepackage{booktabs}                  

\onlineid{1010}

\vgtccategory{Research}

\vgtcinsertpkg

\preprinttext{To appear in 2021 IEEE Visualization \& Visual Analytics (VIS) Poster Program.}

\title{A Visualization Authoring Model for Post-WIMP Interfaces}

\author{
    Marc Satkowski \thanks{e-mail: [msatkowski, weizhouluo, dachselt]@acm.org}\\ 
    \scriptsize Interactive Media Lab\\
    \scriptsize Technische Universit\"at Dresden\\ 
    \and
    Weizhou Luo\footnotemark[1]{}\\
    \scriptsize Interactive Media Lab\\
    \scriptsize Technische Universit\"at Dresden\\ 
    \and
    Raimund Dachselt\footnotemark[1]{} \\
    \scriptsize Interactive Media Lab\\ 
    \scriptsize Technische Universit\"at Dresden\\ 
}

\teaser{
  \centering
  \includegraphics[width=\linewidth]{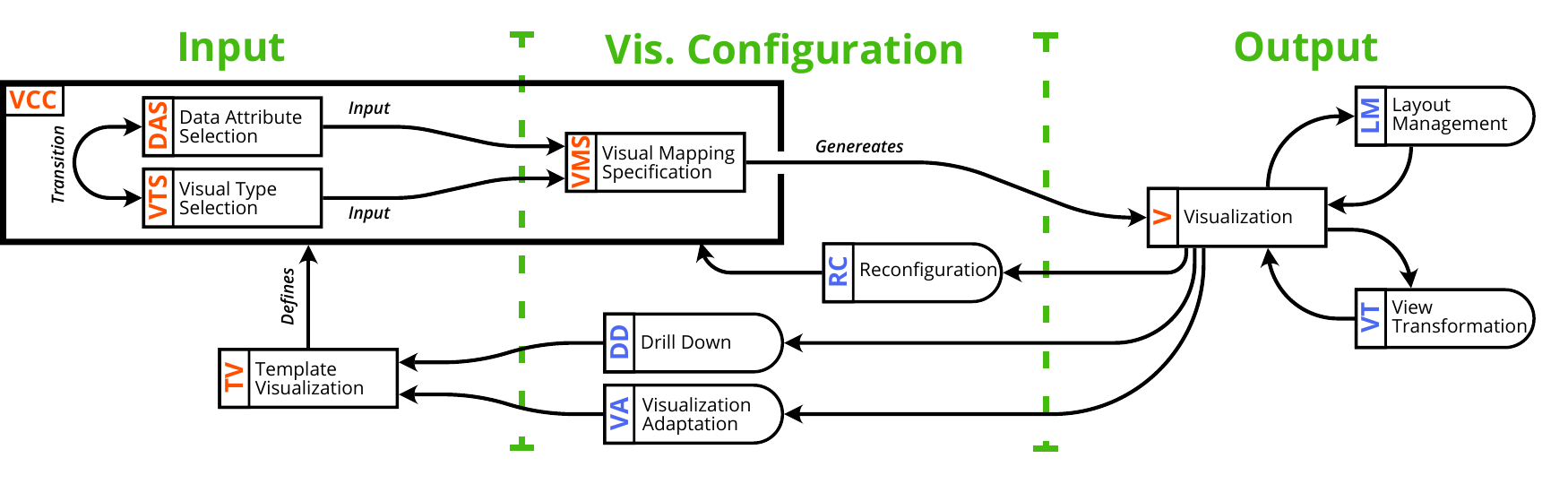}
  \caption{
        Our proposed authoring model.
        It can be split up into three different phases: \textit{Input}, \textit{Visualization Configuration}, and \textit{Output}.
  }
  \label{fig:authoring-model}
}

\abstract{

Besides the ability to utilize visualizations, the process of creating and authoring them is of equal importance.
However, for visualization environments beyond the desktop, like multi-display or immersive analytics environments, this process is often decoupled from the place where the visualization is actually used.
This separation makes it hard for authors, developers, or users of such systems to understand, what consequences different choices they made will have for the created visualizations.
We present an extended visualization authoring model for Post-WIMP interfaces, which support designers by a more seamless approach of developing and utilizing visualizations.
With it, our emphasis is on 
    the iterative nature of creating and configuring visualizations, 
    the existence of multiple views in the same system, and
    requirements for the data analysis process.
}

\CCScatlist{
  \CCScatTwelve{Human-centered computing}{Visu\-al\-iza\-tion}{}{};
  \CCScatTwelve{Human-centered computing}{Visu\-al\-iza\-tion}{Visu\-al\-iza\-tion the\-o\-ry, con\-cepts and pa\-ra\-digms}{}
}

\usepackage{ifthen}
\usepackage{xcolor}

\usepackage{subfigure}
\usepackage{enumitem}
\setlist{nosep}

\usepackage{tabularx}
\usepackage{multirow}
\usepackage{multicol}

\newlength{\tabColSeperator}
\definecolor{NoteBlue}{RGB}{0,0,191}

\definecolor{TodoRed}{RGB}{225,63,63}

\definecolor{ControversialGreen}{RGB}{0,127,0}

\definecolor{UnfinishedBlue}{RGB}{0,127,127}

\definecolor{KKOrange}{RGB}{255,103,2}

\newcommand{\etal}[1]{{#1}~et~al.}
\newcommand{\fref}[2]{Fig.~\ref{#1}#2}

\begin{document}

\maketitle


\section{Introduction}

Information visualizations can be shown on traditional mediums like paper or a desktop monitor, but also in more recent computing environments beyond the desktop \cite{Roberts2014}.
One example of those is Augmented Reality (AR), which enables embedding virtual content into real-world environments.
However, before a visualization can even be presented and thus consumed, explored, or analyzed, it has to be created and configured in the first place, which is also highly important for AR applications \cite{Ashtari2020}.
The example of Immersive Analytics \cite{Marriott2018} shows that it is possible to visualize information quite easily, yet, the construction of visualizations using the same technology is rarely supported.
In most cases, developers and researchers have to create a visualization at their desktop before they can deploy it to a Post-WIMP environment like AR glasses.
This current situation is not only cumbersome, but also leads to an increasing discrepancy between the configuration and presentation process.
Moreover, as demonstrated by the information visualization pipeline \cite{Card1999} or the visualization construction cycle \cite{Grammel2010}, the user is an important part of the whole construction and authoring process.
Therefore, it is necessary to support the user in the Post-WIMP era in creating visualizations by, e.g.,
    overcoming the temporal, spatial, and conceptual distance \cite{Grammel2013} of an authoring application, 
    but also by promoting the general flow \cite{Tominski2020} and 
    the iterative nature \cite{Grammel2010} of the creation process.
In the following, we will present a new authoring model, which 
    builds upon already existing models and research, 
    that is devoted to bridging the gap between visualization authoring and presentation, and
    is applicable not only as a guideline for immersive and other Post-WIMP environments, but also for more common visualization applications.


\section{Constructing and Authoring of Visualizations} 
\label{GeneralWorkflow}

The creation of visualization is an integral part of the data analysis process, and is often labeled as visualization authoring (e.g., \cite{Satyanarayan2020}) or construction (e.g., \cite{Grammel2013,Pantazos2012}).
There already exist models that describe the configuration of visualizations, like the information visualization pipeline of \etal{Card} \cite{Card1999,Chi1998} or the visualization construction cycle of \etal{Grammel} \cite{Grammel2010}.
While in the former the user has the opportunity to influence the creation at different stages, in the latter the user themselves define when to move from one stage to another.
However, visualizations are not created completely individually or separately and data analysis often includes several visualizations at the same time (e.g., Multiple Coordinated Views).
Therefore, the relation or placement of multiple visualizations, but also their meanings with regard to the construction of other visualizations are a necessary factor in the general visualization authoring process.
In the following, we will first discuss possible user and usage perspectives of an authoring process, before we present our
own visualization authoring model (see \fref{fig:authoring-model}), which is inspired by prior work, but also considers a fluid interaction \cite{Tominski2020} between different stages and the existence of several visualizations at the same time.

\subsection{User and Usage Perspective}

One crucial aspect for the authoring of visualizations is the understanding of who is going to create the visualization and for what purposes.
In particular, data analysis is inherently a collaborative activity among various stakeholders with diverse backgrounds, which can result in a collaborative analysis process, or only in a presentation of results and insights in form of visualizations.
This leads to the fact that not only data analysts, but also novice, savvy, or expert users \cite{Pantazos2012} with regard to information visualization have the desire to configure, author, read, and understand visualizations.
Further, it can also be differentiated between the designers or developers and domain users \cite{VanWijk2006,Munzner2014} of visualizations, which vary in the degree of data understanding they have.
In association with different types of users the goal they want to achieve also changes.
Therefore, it is possible to differentiate between data presentation and data exploration \cite{Grammel2013}, or between the wish to produce or to consume visualizations \cite{Munzner2014}.
In general, an authoring model and application should have the goal to enable those diverse usages and different user groups by either supporting their already existing mental model or by helping them developing a new one \cite{Grammel2013}.

\subsection{Visualization Authoring Model}

At the core of our model (see \fref{fig:authoring-model}) resides an altered version of the visualization construction cycle (VCC) of \etal{Grammel} \cite{Grammel2010}.
In there, users of an authoring tool can start by either selecting the data attributes (DAS) to visualize or choosing the visualization type they want to see (VTS).
In the following visual mapping (VMS), the user can define different aspects of a visualization, like 
    the visual encoding \cite{Munzner2014}, 
    the visual marks to use, 
    the data binding, 
    the scale, axes, or legends, and 
    the layout inside the visualization \cite{Satyanarayan2020}.
\par
As a result of the previous operations, a visualization output (V) is produced.
The view of this output can then be further transformed (VT) \cite{Card1999, Chi1998} by different actions the user can take, like panning and zooming, which in return reshapes the visualization output (V).
However, the construction and authoring of a visualization is not only unidirectional but should also allow a user to manipulate different previously made configurations in the authoring pipeline.
Therefore, it should be possible to reconfigure (RC) different parts of the visualization construction cycle (VCC).
However, this reconfiguration alters and therefore destroys the currently existing visual appearance of the visualization.
To be able to construct an alternative look of this view (VA), it is possible to create a new visualization based on the already existing one.
Additionally, an existing visualization can also be used as a basis for further 
    aggregations, 
    refinements, or 
    drill downs (DD), 
based on the data it displays or is selected \cite{Guimaraes2011}.
In general, both mentioned approaches use another visualization as a template (TV), which then can be used to predefine different values in the visualization construction cycle (VCC).
\par
Lastly, as the data analysis process often involves multiple visualizations, it is also possible to not only create one, but several visualizations at the same time.
This makes it necessary to allow for different layout management behaviours (LM) to structure the presentation of arbitrary groups of visualizations and help with the later sensemaking process.
Those layout adaptations could include, e.g., 
    the position of visualizations or
    uniform axis between each view.
\par
Our authoring model can further be split up into three phases focused on 
    the \textit{input}, 
    the \textit{visualization configuration}, and 
    the \textit{output}.
In \textit{Input}, users can start from scratch by selecting the data (DAS) or visualization type (VTS), or can use an already created visualization in the environment which predefines several parameters (TV).
This follows the notion of \etal{Grammel} \cite{Grammel2013}, which, in contrast to \etal{Card}'s \cite{Card1999} information visualization pipeline, lists two types of visual mapping approaches: data-driven and visualization-driven.
In \textit{Visualization Configuration}, the user can specify the visual mapping (VMS) of the data and visualization type.
Further, they
    can change already chosen values (RC) of an existing visualization, 
    use the same for the creation of a new visualization which alters visual or data attributes (VA), or 
    use the visualization as a drill down (DD) starting point.
In \textit{Output}, the created visualization (V) can be transformed (VT) by the user, or, in cases with more than one view, manage the layout (LM) those visualizations are presented in.

\section{Conclusion}

In our work we presented an extended visualization authoring model. 
With this we showed the high level of interconnection of a visualization, the configuration process, and its input, which makes it necessary to bring the three phases of the model as closely together as possible for future authoring applications.
Further, we believe that this model can be applied to any visualization configuration system, but is particularly important for novel beyond the desktop applications.
To prove this, we already use our model in our current research project which revolves around designing and implementing an AR authoring system usable for immersive analytics environments.


\acknowledgments{    
    This work was funded in part by ``Deutsche Forschungsgemeinschaft'' (DFG, German Research Foundation) 
    under grant number 319919706/RTG2323 ``Conducive Design of Cyber-Physical Production Systems'', and 
    under project number 389792660 as part of TRR~248 (see \url{https://perspicuous-computing.science}).
}

\bibliographystyle{abbrv-doi}

\bibliography{main}
\end{document}